\documentclass[12pt,british,british]{article}
\usepackage[T1]{fontenc}
\usepackage[latin1]{inputenc}
\usepackage{a4wide}
\usepackage{babel}
\usepackage{graphics}

\makeatletter

\providecommand{\LyX}{L\kern-.1667em\lower.25em\hbox{Y}\kern-.125emX\@}

\usepackage[T1]{fontenc}
\usepackage[latin1]{inputenc}
\usepackage{a4}
\usepackage{babel}


\begin{document}
{\raggedleft July 2001\par}

{\raggedleft ULG-PNT-JRC-2001-1\par}
\bigskip{}

\vspace{1.5cm}
{\centering {\large Does \( F_{2} \) need a hard pomeron?}\large \par}

\bigskip{}
{\centering J.R. Cudell\footnote{%
JR.Cudell@ulg.ac.be 
} and G. Soyez\footnote{%
G.Soyez@ulg.ac.be 
} \par}

{\centering {\small Inst. de Physique, Bât. B5, Université de Liège,
Sart Tilman, B4000 Liège, Belgium\par{}}\small \par}
\vspace{1.5cm}

\begin{abstract}
\noindent We show that the latest HERA measurements of \( F_{2} \)
are compatible with the existence of a triple pole at \( J=1 \).
Such a structure also accounts for soft forward hadronic amplitudes,
so that the introduction of a new singularity is not necessary to
describe DIS data.\bigskip{}
\end{abstract}
The latest data from HERA \cite{Adloff:2000qk, Breitweg:2000yn, H1highQ}
have attained such a level of precision that they can constrain quite
stringently the possible singularity structures in the complex \( J \)
plane which control the \( s \) dependence of soft hadronic amplitudes
at \( t=0 \) and the \( x \) dependence of \( F_{2} \). Traditionally,
it has been assumed that soft hadronic amplitudes would be constrained
by their analytic properties, \emph{i.e.} by Regge theory, whereas
DIS would be the domain where perturbative QCD could be tested, with
an overlap at only extremely small \( x \). However, it has recently
been realised \cite{Donnachie:1998gm, Cudell:1999kf} that there is
no reason to limit the application of Regge theory to such a domain.
Indeed, hadronic scattering amplitudes are the continuation of \( t \)-channel
scattering amplitudes, and the simple structures predicted by Regge
theory become valid once the cosine of the \( t \)-channel scattering
angle, \( \cos \theta _{t} \), becomes sufficiently large. One can
in fact quantify this in the hadron-hadron case, as the forward scattering
amplitudes (measured through \( \rho  \) and \( \sigma _{tot} \))
are well fitted for \( \sqrt{s}>10 \) GeV. This corresponds, in the
\( pp \) and \( \overline{p}p \) cases, to a value of 

\begin{equation}
\label{cos}
\cos \theta _{t}=\frac{s}{2m_{p}^{2}}\geq 57.
\end{equation}
 In the \( \gamma ^{*}p \) case, the total cross section, \begin{equation}
\label{sigma}
\sigma _{\gamma ^{*}p}=\frac{4\pi ^{2}\alpha _{elm}}{Q^{2}}F_{2}(x,Q^{2}),
\end{equation}
 should be described by the same singularities for the same region
of \( \cos \theta _{t}=\nu /(m_{p}\sqrt{Q^{2}})=\sqrt{Q^{2}}/(2m_{p}x). \)
Hence one would expect to observe the same singularities in \( F_{2} \)
and in total hadronic cross sections for values

\begin{equation}
\label{xrange}
x\leq \frac{\sqrt{Q^{2}}}{100m_{p}}.
\end{equation}

This is a rather large interval where there is considerable overlap
with standard perturbative QCD evolution, hence one would hope for
the two descriptions to become compatible (but we are not yet at a
stage to show this result). Whether it makes sense to extend Regge
fits to larger \( x \) \cite{Donnachie:2001xx, Desgrolard:2001}
or to smaller \( s \) \cite{Desgrolard:2001} remains debatable.

The problem in implementing such a program is that, although the same
singularities have to be present, their residues are not known, and
must be \( Q^{2} \)-dependent. It is known however that the \( Q^{2} \)
dependence must enter only in the residues, \emph{i.e.} the \( \nu  \)
dependence should not change with \( Q^{2} \). As the effective \( \nu  \)
dependence does seem to vary as \( Q^{2} \) increases at HERA, this
can only be obtained by combining several singularities, and writing

\begin{equation}
\label{F2regge}
F_{2}(\frac{Q^{2}}{2\nu },Q^{2})=\sum _{i}g_{i}(Q^{2})f_{i}(\nu )
\end{equation}
with the \( f_{i} \) resulting from the singularities observed in
fits to total cross sections \cite{PDG2000}. The fact that the residues
depend on \( Q^{2} \) means that some singularities may remain hidden
in soft scattering amplitudes, and manifest themselves only at high
\( Q^{2} \). However, as the values of \( Q^{2} \) observed now
at HERA span the whole range from 0.045 to 30000 GeV\( ^{2} \), it
seems unlikely that such singularities would switch off totally in
total cross sections\footnote{%
It is however also possible to argue that the photon is special, and
that t-channel unitarity relations, from which the universality of
residues is derived \cite{Gribov}, do not form a closed system for
photons, hence they could have exceptional singularities. 
}. 

The first model proposing such a global fit is due to Donnachie and
Landshoff \cite{Donnachie:1998gm} and is based on their previous
model \cite{Donnachie:1986iz} based on a simple-pole singularity,
which accounts very successfully not only for hadronic total cross
sections, but also for elastic and diffractive scattering \cite{Donnachie:1986iz}.
The problem is that, at small \( x \), this model predicts a single
singularity in (\ref{F2regge}), and hence the effective power of
\( x \) cannot change with \( Q^{2} \). This lead Donnachie and
Landshoff to postulate the existence of a hidden singularity, called
the hard pomeron, which is needed to reproduce the rise of \( F_{2} \)
at small \( x \), but which is totally absent from total hadronic
cross sections, with residues at least \( 10^{6} \) times smaller
than those of the soft pomeron \cite{PDG2000}. It is rather hard
to imagine how such a fierce singularity can totally turn off in hadron-hadron
scattering, as the hadrons will sometimes fluctuate into small systems
of quarks, comparable to those produced by the off-shell photons.
Furthermore, the model seems to require the soft pomeron to turn off
in \( F_{2}^{c} \), whereas it is needed in \( J/\psi  \) production.
Although one cannot prove that these features are excluded, they seem
rather unnatural. Nevertheless, the model reproduces quite well the
new data from HERA, and can even be extended outside the area of applicability
of Regge theory \cite{Donnachie:2001xx}.

There are other alternative fits to total cross sections \cite{Desgrolard:2001, Gauron:2000}
(for a review see \cite{PDG2000, COMPETE}) which work equally well,
or better, than simple poles. In such models, the rise of total cross
sections is assumed to come from a double pole, or a triple pole,
at \( J=1 \).

In the first case \cite{Desgrolard:2001}, the fit to soft amplitudes
forces one to assume a large splitting of the intercepts of leading
meson trajectories, which means that such trajectories must be non
linear. Furthermore, the leading \( C=+1 \) contribution becomes
negative at sufficiently small \( s \), of the order of 10 GeV, and
factorisation would naively lead to negative total cross sections
for processes which couple only to the leading trajectories. Although
once again these points cannot be used to rule the model out (as neither
factorisation of double poles nor the linearity of meson trajectories
can be proven), they make it seem unnatural.

The second candidate \cite{Gauron:2000} does not have these drawbacks,
and the description of total cross sections that it provides can even
be extended down to \( \sqrt{s}=5 \) GeV \cite{COMPETE}. Not only
does it fit well both total cross sections, and values of the \( \rho  \)
parameter for forward hadronic amplitudes, but the outcome of such
a fit seems the most natural: unitarity is preserved, the various
\( C=+1 \) terms of the cross section remain positive, and the meson
trajectories are compatible with exchange degeneracy. We shall show
here that this fit extends naturally to reproduce DIS data, at least
in the Regge region (\ref{xrange}).

We shall adopt here the natural Regge variable \( 2\nu  \) instead
of \( W \) \cite{Desgrolard:2001}, as this makes a sizable difference
in the large-\( x \) region, and we shall assume that \( F_{2} \)
can be fitted by

\begin{equation}
\label{functionalform}
F_{2}(\frac{Q^{2}}{2\nu },Q^{2})=a(Q^{2})\log ^{2}\left( \frac{2\nu }{2\nu _{0}(Q^{2})}\right) +c(Q^{2})+d(Q^{2})(2\nu )^{-0.47}
\end{equation}
in the region (\ref{xrange}). The first two terms represent pomeron
exchange, and the third \( a/f \) exchange. We have fixed the intercept
of the latter from fits to hadronic amplitudes \cite{PDG2000, COMPETE}. 

The first test to check whether such a simple form has a chance to
reproduce the data is to follow \cite{Donnachie:1998gm} and extract
the residues \( a(Q^{2}) \), \( c(Q^{2}) \), \( d(Q^{2}) \) and
the scale \( b(Q^{2})\equiv \log \left( 2\nu _{0}(Q^{2})\right)  \)
by fitting the \( \nu  \) dependence for each value of \( Q^{2} \). 

As explained above, we limit ourselves to a consideration of the data
in the region (\ref{xrange}), and to \( Q^{2} \) bins where there
are at least 10 data points. We show in Fig. 1 the result of such
a fit (to a total of 875 points), which corresponds to having one
parameter for each residue and for each value of \( Q^{2} \). This
fit has a \( \chi ^{2} \) per data point of 0.674, compared with
1.00 for the hard pomeron model of \cite{Donnachie:1998gm} refitted
to those data. Hence it seems that \emph{a priori} the data are as
compatible with (\ref{functionalform}) as with the fits of \cite{Donnachie:1998gm, Donnachie:2001xx}.
The graphs represent the coefficients respectively of the \( \log ^{2} \)(2\( \nu  \))
, \( \log  \)(2\( \nu  \)) and constant terms. We see that the data
do exhibit a suppression at small \( Q^{2}, \) and there is a strong
correlation between the three terms, which all seem to have a similar
behaviour with \( Q^{2} \). 

\bigskip{}
{\centering {\footnotesize (a)~~~~~~~~~~~~~~~~~~~~~~~~~~~~~~~~~~~~~~~~~~(b)~~~~}\footnotesize \par}

\vglue -2cm

\vspace*{-0cm}
{\centering \resizebox*{!}{7cm}{\includegraphics{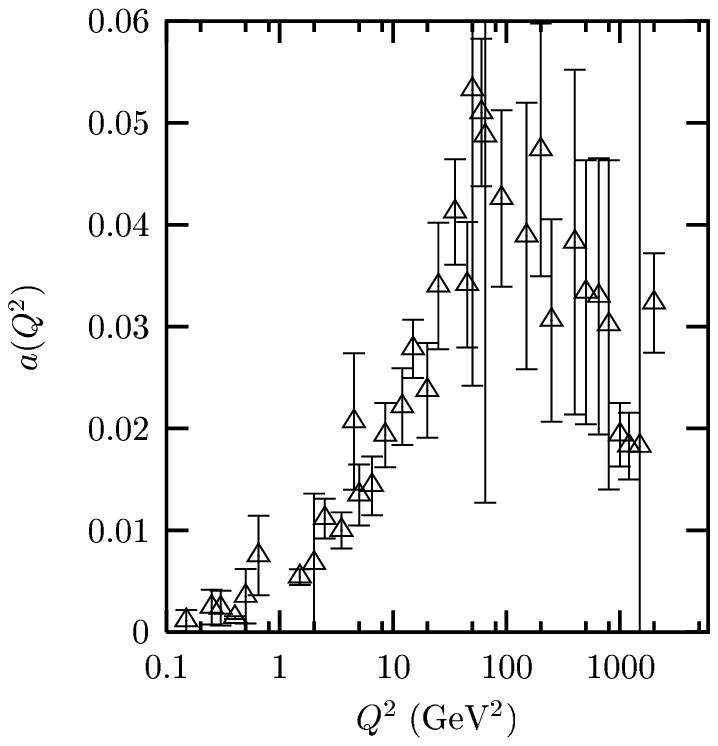}}  \resizebox*{!}{7cm}{\includegraphics{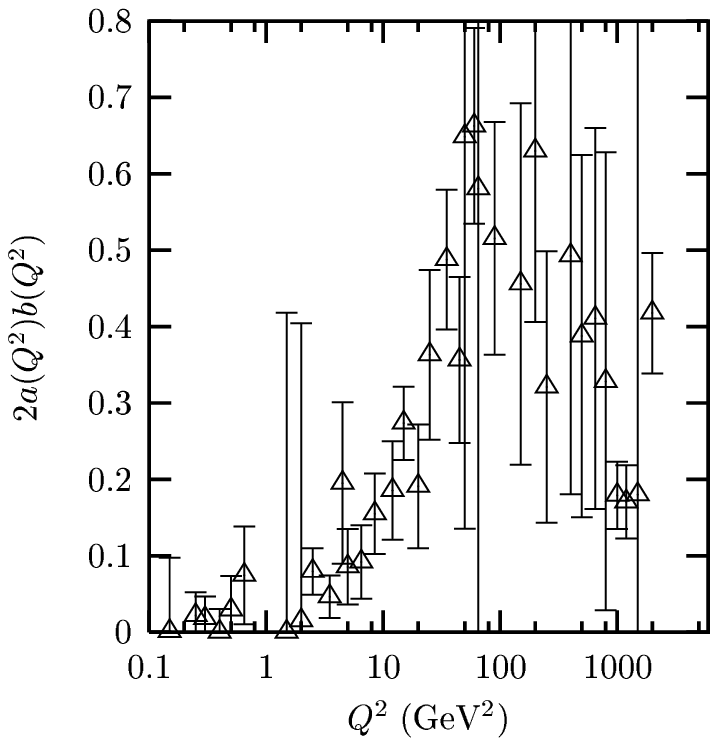}} \par}

{\centering \vglue 2cm\par}
\vspace{0.3cm}

\bigskip{}
{\centering {\footnotesize (c)}\footnotesize \par}

\vglue -2cm

{\centering \resizebox*{!}{7cm}{\includegraphics{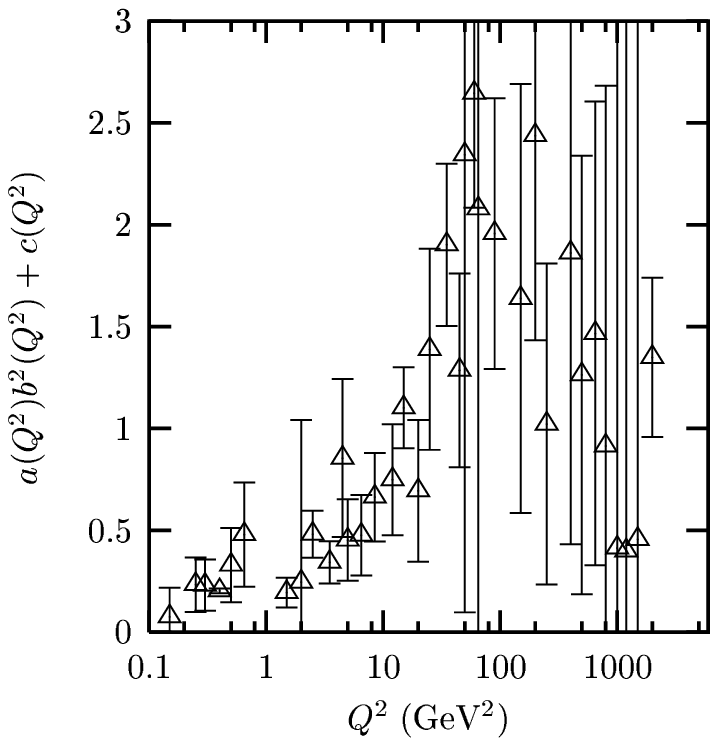}} \par}
\vspace{0.3cm}

\begin{quote}
{\small Figure 1: the coefficients (a) of log\( ^{2}(2v) \), (b)
of log\( (2\nu ) \) and (c) of the constant term, extracted from
the latest HERA data.}{\small \par}
\end{quote}
We can now proceed to a parametrisation of the residues. We first
consider the data of refs. \cite{Adloff:2000qk, Breitweg:2000yn},
which extend to \( Q^{2}=135 \) GeV\( ^{2} \), as they are almost
entirely in region (\ref{xrange}), together with data for the total
\( \gamma p \) cross section \cite{gammadata} at \( \sqrt{s}> \)5
GeV\footnote{%
As the data from 5 to 10 GeV have big error bars, it makes little
difference whether we include them or not.
}. Apart from the fact that gauge invariance imposes that the form
factors vanish linearly at \( Q^{2}\rightarrow 0 \) for fixed \( \nu  \),
their form is undetermined. We find that a good fit is obtained by
the following forms: \begin{equation}
\label{Eq1a}
g(Q^{2})=A_{g}Q^{2}\left( \frac{1}{1+Q^{2}/Q_{g}^{2}}\right) ^{\epsilon _{g}},\: g=a,\, c,\, d
\end{equation}
 \begin{equation}
\label{Eq1b}
b(Q^{2})=\log \left[ 2\nu _{0}(Q^{2})\right] =b_{0}+b_{1}\left( \frac{Q^{2}}{Q^{2}+Q_{b}^{2}}\right) ^{\epsilon _{_{b}}}
\end{equation}

\bigskip{}

{\centering \begin{tabular}{|c|c|c||c|c|c|}
\hline 
\multicolumn{3}{|c||}{\centering \( Q^{2}\leq 135 \) GeV\( ^{2} \)}&
\multicolumn{3}{|c|}{\( Q^{2}\leq 30000 \) GeV\( ^{2} \)}\\
\hline
\hline 
{\small parameter}&
{\small value}&
{\small error}&
{\small parameter}&
{\small value}&
{\small error}\\
\hline 
{\small \( A_{a} \)}&
{\small 0.00981}&
{\small 0.0031}&
{\small \( A_{a} \)}&
.99383E-02&
.17197E-03\\
\hline 
{\small \( Q_{a} \)}&
{\small 0.992}&
{\small 0.125}&
{\small \( Q_{a} \)}&
1.8850&
.74726E-01\\
\hline 
{\small \( \epsilon _{a} \)}&
{\small 0.721}&
{\small 0.026}&
{\small \( \epsilon _{a} \)}&
.89988&
.10849E-01\\
\hline 
{\small \( A_{c} \)}&
{\small 0.945}&
{\small 0.008}&
{\small \( A_{c} \)}&
.95727&
.38888E-02~\\
\hline 
{\small \( Q_{c} \)}&
{\small 0.696}&
{\small 0.035}&
{\small \( Q_{c} \)}&
.62391&
.16433E-01\\
\hline 
{\small \( \epsilon _{c} \)}&
{\small 1.34}&
{\small 0.04}&
{\small \( \epsilon _{c} \)}&
1.3903&
.23285E-01\\
\hline 
{\small \( A_{d} \)}&
{\small 0.430}&
{\small 0.066}&
{\small \( A_{d} \)}&
.27401&
.27215E-01\\
\hline 
{\small \( Q_{d} \)}&
{\small 0.260}&
{\small 0.367}&
{\small \( Q_{d} \)}&
31.987&
5.5734\\
\hline 
\( \epsilon _{d} \)&
{\small 0.451}&
{\small 0.104}&
\( \epsilon _{d} \)&
1.6884&
.15020\\
\hline 
\( b_{0} \)&
{\small 3.00}&
{\small 0.641}&
\( b^{'}_{0} \)&
3.00&
.91789E-02\\
\hline 
\( b_{1} \)&
{\small 3.31}&
{\small 0.27}&
\( b^{'}_{1} \)&
.13797&
.54106E-01\\
\hline 
\( Q_{b} \)&
{\small 18.2}&
{\small 8.92}&
\( Q^{'}_{b} \)&
4.6269&
1.6797\\
\hline 
\( \epsilon _{b} \)&
{\small 3.16}&
{\small 1.26}&
\( \epsilon ^{'}_{b} \)&
1.8551&
.15402\\
\hline
\end{tabular}\par}

\begin{quote}
\vspace*{\medskipamount}
{\small Table 1: the values of the parameters corresponding to Eqs.
(\ref{Eq1a}, \ref{Eq1b}) for the low-to-intermediate \( Q^{2} \)
fit, and those of the global fit to region (\ref{xrange}), corresponding
to Eqs. (\ref{Eq1a},\ref{newb}).}{\small \par}
\bigskip{}

{\centering \includegraphics{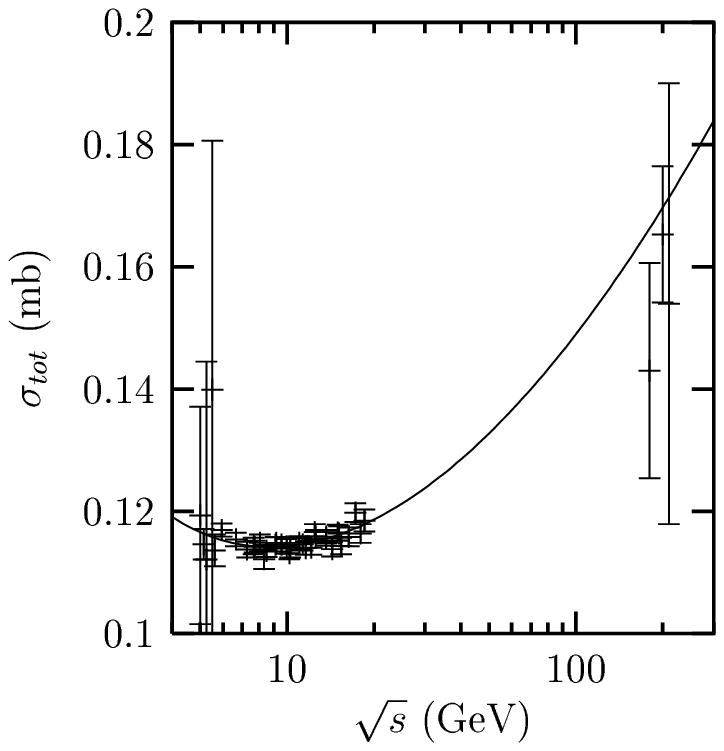} \par}
\end{quote}
{\centering {\small Figure 2: fit to the total cross section data
for \( \sqrt{s}\geq 5 \) GeV. }\small \par}
\vspace{0.3cm}

\begin{quote}
The resulting values of the parameters are given in Table 1. We imposed
that the fit smoothly reproduces the value of \( \nu _{0} \)(0) which
results from a global fit to all hadronic cross sections \cite{PDG2000}.
The fit gives a \( \chi ^{2} \) of 185.2 for 241 points (including
38 points for the total cross section), which is somewhat better than
those of \cite{Donnachie:2001xx, Desgrolard:2001}. We show in Figs.
2 and 3 the curves corresponding to these results.
\end{quote}
\vspace{0.3cm}
{\centering \resizebox*{10cm}{!}{\includegraphics{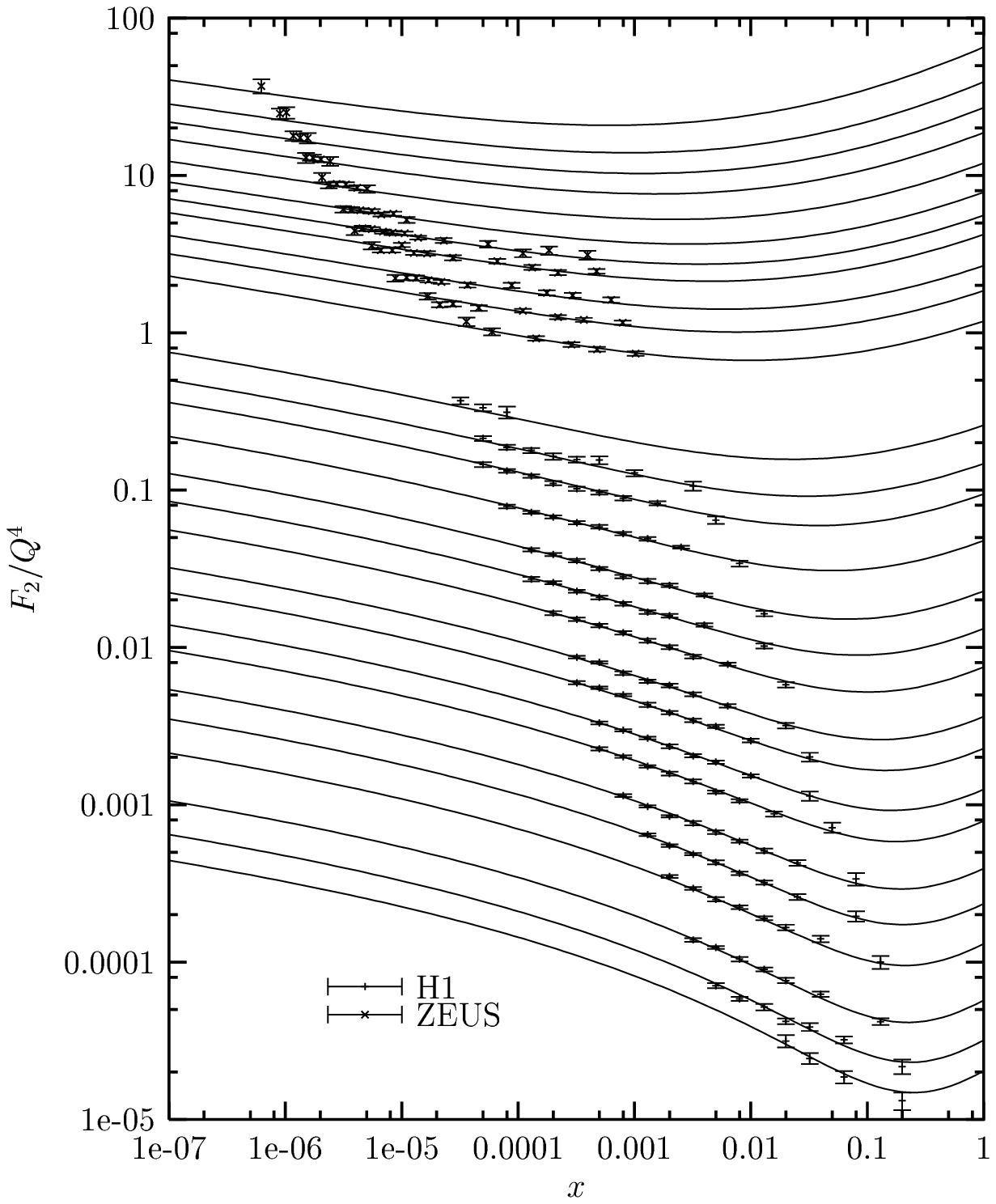}} \par}
\vspace{0.3cm}

\begin{quote}
{\centering {\small Figure 3: fit to the new HERA data \cite{Adloff:2000qk, Breitweg:2000yn}
for \( Q^{2}\leq 135 \) GeV\( ^{2} \). }\small \par}
\end{quote}
It is interesting to note that such simple forms for the residues,
which achieve a remarkably low \( \chi ^{2} \) up to 135 GeV\( ^{2} \),
do not extend well beyond 500 GeV\( ^{2} \). Whether one wants to
go beyond this result is really a matter of taste. As we have explained,
it is possible to have {}``hidden{}'' singularities, which would
manifest themselves only as \( Q^{2} \) becomes large enough. From
a perturbative QCD point of view, the present fit can be seen as a
starting point for \( F_{2}(x,Q^{2}) \), consisting of constants,
\( \log x \) and \( \log ^{2}x \) terms.\footnote{%
One must note however that we have not tried to go beyond the Regge
region, and hence the limit \( x\rightarrow 1 \) is not correct,
as \( F_{2} \) should go to 0 there.
} Perturbative evolution will then produce further \( \log x \) terms,
which would be hidden below the starting evolution scale. As singularities
of hadronic amplitudes cannot occur at arbitrary places, the latter
would have to be a physical parameter, and the present fit shows that
it should be of order \( Q_{0}=10 \) GeV \cite{Capella}. 

It may be a worthwhile exercise however to check whether this singularity
structure can be extended to the full Regge region, and to consider
a fit to the full dataset of DIS \cite{datsetDIS}. The reason is
that in this case there is considerable overlap between the DGLAP
region (\( Q^{2} \) large, \( \log Q^{2}\gg -\log x \)) and the
Regge region (\ref{xrange}), and hence one should be able to understand
some features of the pomeron through perturbative means. In this region,
to achieve values of \( \chi ^{2} \) comparable to those of \cite{Donnachie:2001xx, Desgrolard:2001},
we need in fact to modify the form used for the scale \( \nu _{0}(Q^{2}) \),
and introduce a logarithm in its expression:

\begin{equation}
\label{newb}
b(Q^{2})=b_{0}+b_{1}^{'}\left[ \log \left( 1+\frac{Q^{2}}{Q_{b}^{'\, 2}}\right) \right] ^{\epsilon '_{b}}.
\end{equation}
Such a form allows us to extend the fit to the full Regge region (\ref{xrange}),
and produces a reasonable \( \chi ^{2}/dof \): we obtain 1411 for
1166 points (including 21 points for the total cross section above
\( \sqrt{s}=10 \) GeV)\footnote{%
Note that the fit of \cite{Donnachie:2001xx} give a \( \chi ^{2} \)
of 3941 on those points.
}, but the \( \chi ^{2} \) for the new HERA points \cite{Adloff:2000qk, Breitweg:2000yn}
gets degraded to 311. This is largely due to the tiny size of the
errors on the new data, and to some inconsistencies in the full dataset.
A fine-tuning of the form factors \cite{Desgrolard:2001} could presumably
lead to a better \( \chi ^{2} \), but what we would learn from such
an exercise is unclear. We show in Fig. 4 the results of this global
fit, and, as we can see, the full Regge region (\ref{xrange}) is
well accounted for. Note that although all the data are fitted to,
we show only the data from HERA, as the number of values of \( Q^{2} \)
would otherwise be too large to represent in this manner.

In conclusion, we see that several scenarios compatible with Regge
theory are possible to describe structure functions. All of them share
the characteristic that one needs several component to describe DIS
and soft scattering at the same time. We believe that the present
parametrisation shows that no unexpected behaviour is needed to reproduce
DIS, and that the pomeron may well be a single object, connecting
all regions of \( Q^{2} \) smoothly, and exhibiting the same singularities
in DIS and in soft scattering. How to obtain such a simple form in
the region of overlap between perturbative QCD and Regge theory remains
an open question.

\section*{Acknowledgements}

We thank P.V. Landshoff for fruitful exchanges, and E. Martynov for
sharing his results with us, for numerous discussions, and for correcting
a mistake in a preliminary version.

\vspace{0.3cm}
{\centering \resizebox*{0.9\textwidth}{!}{\includegraphics{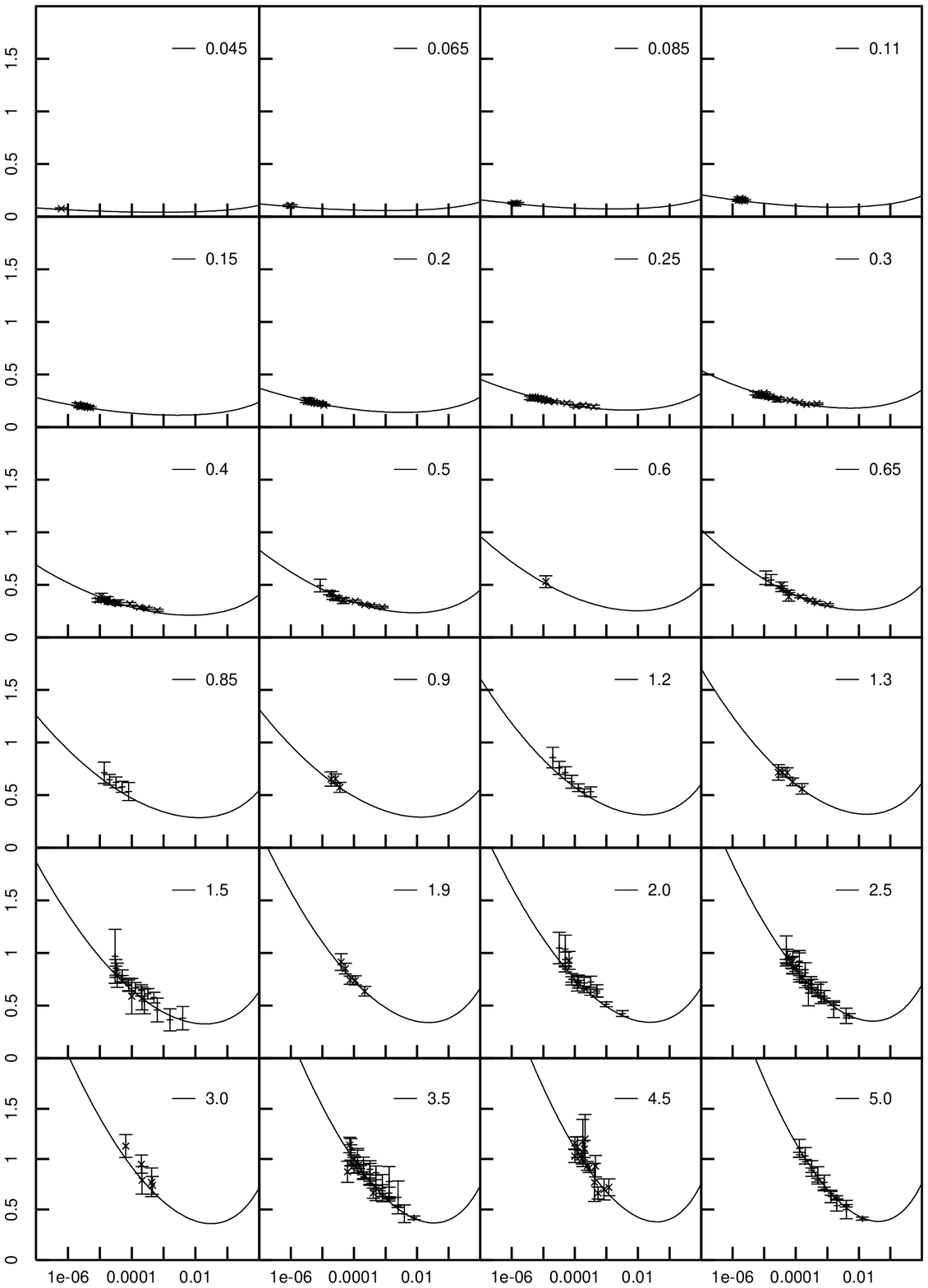}} \par}

\begin{quote}
{\small \vglue -2cm Figure 4 (a): Result of a global fit to all the
data in the Regge region, for \( 0.045\leq Q^{2}\leq 5.0 \) GeV\( ^{2} \).
\( F_{2}(x,Q^{2}) \) is shown as a function of \( x \), for each
\( Q^{2} \) value indicated (in GeV\( ^{2} \)). Only data from HERA
are shown.}{\small \par}
\end{quote}
\vspace{0.3cm}
{\centering \resizebox*{0.9\textwidth}{!}{\includegraphics{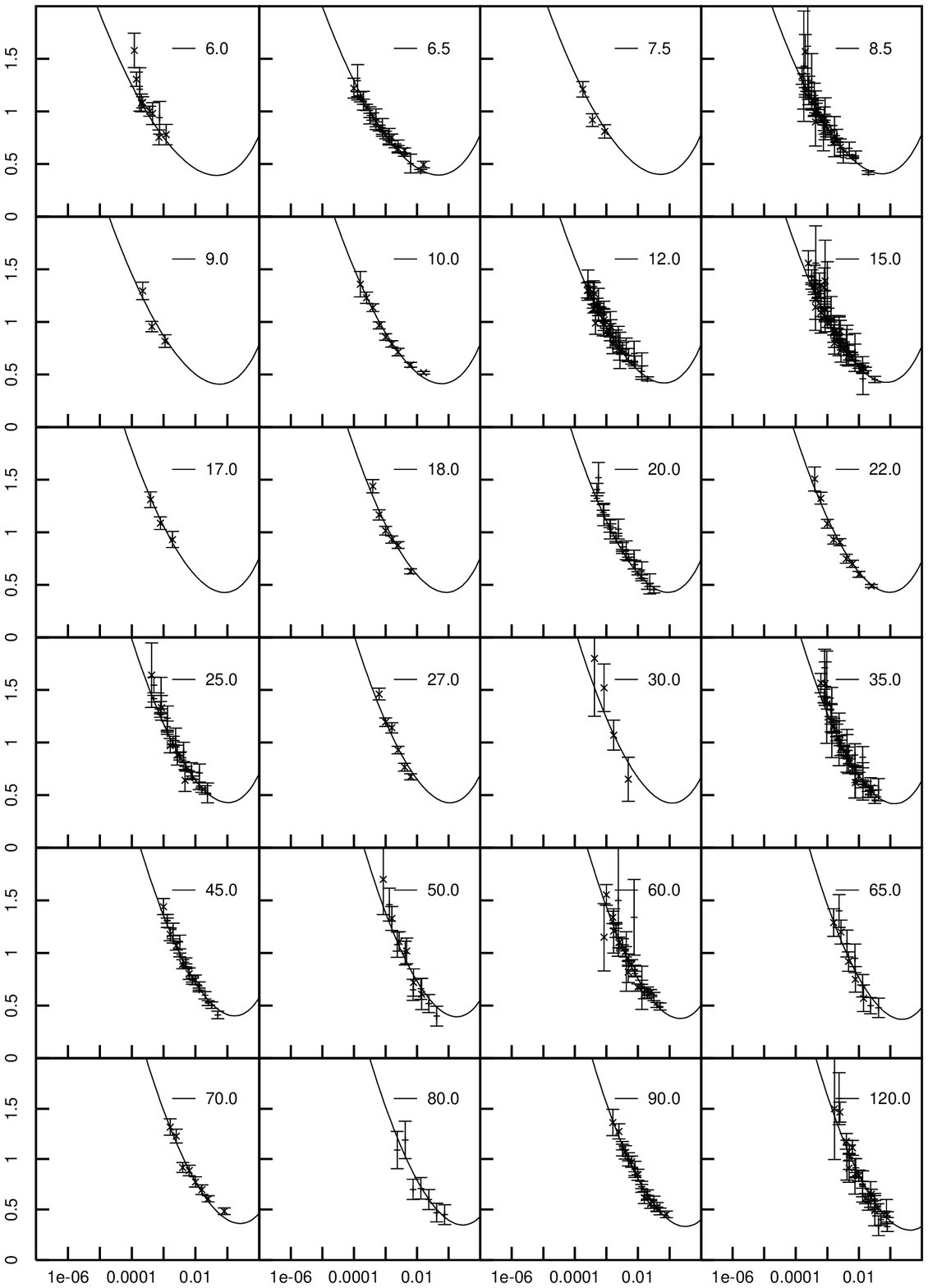}} \par}
\vspace{0.3cm}

\begin{quote}
{\small \vglue -2cm Figure 4 (b): Result of a global fit to all the
data in the Regge region, for \( 6.0\leq Q^{2}\leq 120.0 \) GeV\( ^{2} \).
\( F_{2}(x,Q^{2}) \) is shown as a function of \( x \), for each
\( Q^{2} \) value indicated (in GeV\( ^{2} \)). Only data from HERA
are shown.}{\small \par}
\end{quote}
{\centering \resizebox*{0.9\textwidth}{!}{\includegraphics{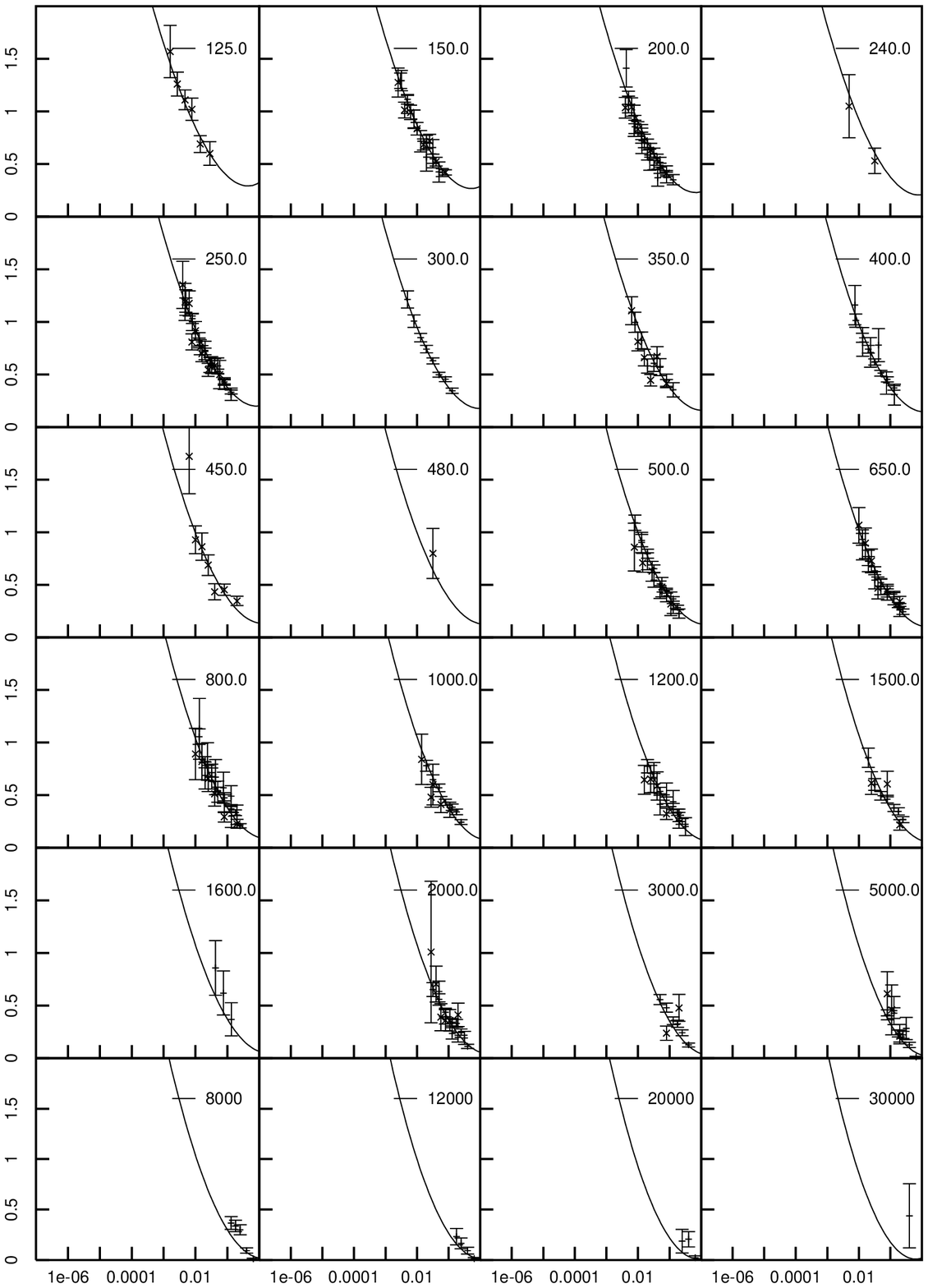}} \par}
\vspace{0.3cm}

\begin{quote}
{\small \vglue -2cm Figure 4 (c): Result of a global fit to all the
data in the Regge region, for \( 125.0\leq Q^{2}\leq 30000.0 \) GeV\( ^{2} \).
\( F_{2}(x,Q^{2}) \) is shown as a function of \( x \), for each
\( Q^{2} \) value indicated (in GeV\( ^{2} \)). Only data from HERA
are shown.}{\small \par}
\end{quote}

~
\end{document}